\documentclass[twocolumn,showpacs]{revtex4}
\usepackage{psfig}
\begin{document}
%\draft
\title{The reaction ${\pi}N{\to}{\omega}N$ revisited: the $\omega{-}N$ scattering
length}

\author{C. Hanhart, A. Sibirtsev, and J. Speth}
\affiliation{$^a$Institut f\"ur Kernphysik, Forschungszentrum J\"ulich, 
D-52425 J\"ulich}
\date{\today}

{FZJ--IKP--TH--2001--13}

\begin{abstract}
We reinvestigate the experimental data on $\omega$--meson production in pion-nucleon 
collisions~\cite{Binnie,Keyne,Karami} based on both an analytical approach and
a Monte Carlo simulation. Our analysis allows us to
study the kinematical peculiarities of the 
$\pi^-p{\to}\omega{n}$ reaction at energies close to threshold.
Based on the hypothesis, that the formalism derived in 
ref. \cite{Binnie} was applied improperly,
 the unusually strong 
energy dependence claimed
for the near--threshold cross section is identified as a purely
kinematical effect. Based on this hypothesis,  we deduce an effective 
${\pi^-}p{\to}{\omega}n$ cross section that is larger by an order of magnitude compared 
to the one commonly used. 
In addition we  extract
a lower limit for the imaginary part $\Im a_{{\omega}N}$ of the ${\omega}N$ 
scattering length. We find a value of $\Im a_{{\omega}N}{=}0.24{\pm}0.05$ 
fm, which is  significantly larger than that presently found in the literature.
\end{abstract}


\maketitle

%%%%%%%%%%%%%%%%%%%%%%%%%%%%%%%%%%%%%%%%%%%%%%%%%%%%%%%%%%%%%%%%%%%%%%%%%%

\section{Introduction}
\label{sec:intro}

In Refs.~\cite{Binnie,Keyne,Karami} the reaction $\pi^-p \to \omega n$ was
measured at energies close to the nominal $\omega$ production threshold.
In Ref.~\cite{Hanhart2} it was claimed that these experiments were interpreted
incorrectly. In this work we reinvestigate
 the experiments using an event--by--event simulation as well as
an analytical calculation. The purpose of this manuscript is two fold: Besides
studying, if the reaction kinematics was treated properly in Refs. 
\cite{Binnie,Keyne,Karami}, we also take the oportunity to discuss in detail
the method proposed in 
Ref. \cite{Binnie} for the near threshold production of narrow resonances, for
it might prove usefull for upcoming experiments at modern accelerators.

The reaction $\pi{N}{\to}\omega{N}$,  especially
at energies close to the reaction threshold, attracts great
interest for several reasons.

1) There is a large number of  baryonic
resonances that have been predicted theoretically but not yet observed 
experimentally~\cite{Isgur1}. Since almost all our knowledge about resonances 
 is deduced from elastic ${\pi}N$ scattering, to really resolve the
issue of the missing resonances other channels must be 
studied. There are predictions of   a number 
of baryonic resonances with masses from 1.8 to 2 GeV which might 
couple to the ${\omega}N$ channel~\cite{Capstick}.
The $\pi{N}{\to}\omega{N}$ data is an essential ingredient
in any kind of analysis aimed at identifying baryonic resonances 
possibly coupled to the 
$\omega$--meson. 
A reliable extraction of the resonance properties 
implies the analysis of pion and photon induced data simultaneously for 
as many final states as possible~\cite{Feuster1,Feuster2}. 

In the experiments described in Refs. \cite{Binnie,Keyne,Karami}
the squared matrix element $|{\cal M}|^2$ was extracted from
a measurement of the reaction $\pi^- p \to \omega n$.
The authors claim to have measured a very strong 
near--threshold suppression, which might stem from the production of a baryonic 
resonance in a $p$--wave. The measured differential cross
sections, however,  are  isotropic.
Up to now a suppression of the matrix element as reported in \cite{Binnie,Keyne,Karami}
cannot be understood theoretically \cite{Wil}.
In any case, the $\pi^-p{\to}\omega{n}$ cross section is very 
important for resolving the existence of resonances not yet observed.

2) The $\pi^-p{\to}\omega{n}$ data~\cite{Karami} are intensively 
used in the evaluation
of the in-medium properties the of $\omega$--meson~\cite{Friman,Lutz1,Klingl1,Lykasov}.
 The in-medium
$\omega$--meson mass and width are commonly considered either for
low finite $\omega$ momenta $p_\omega$ or at $p_\omega$=0.  
Within the $t\rho$ approximation an additional in-medium collisional 
width $\Delta\Gamma_\omega$ of the $\omega$--meson 
is given by the imaginary part of the forward $\omega{N}$ scattering
amplitude ${\Im}f_\omega(0)$ as~\cite{Lenz,Dover,Lutz2},
\begin{equation}
\Delta\Gamma_\omega =4\pi \, \frac{m_N+m_\omega}{m_N m_\omega}
\, \Im f_\omega(0) \, \rho_B,
\end{equation}
where $m_N$ and $m_\omega$ are the free nucleon and $\omega$--meson
masses, respectively, while $\rho_B$ is the baryon density. 
${\Im}f_\omega(0)$ can be evaluated 
using the optical theorem from the total ${\omega}N$ cross section
$\sigma_{{\omega}N}^{tot}$ as
\begin{equation}
\Im f_\omega(0)= \frac{p_\omega}{4\pi}\, \sigma_{{\omega}N}^{tot},
\end{equation}
with $p_\omega$ denoting the $\omega$--meson momentum. 

The partial ${\omega}N{\to}{\pi}N$ cross section can be obtained 
 from the ${\pi}N{\to}{\omega}N$ data using detailed balance.
The total ${\omega}N$ cross section consists of several
different final channels, but the ${\omega}N{\to}{\pi}N$ 
channel is believed to be the one numerically most important.
 In any case, knowledge of this inelastic channel
 provides at least a lower bound
for the imaginary part of the forward scattering amplitude
${\Im}f_\omega(0)$. 
The above procedure was extensively applied in the literature~\cite{Friman,Lutz1}.

3) The feasibility of producing nuclear bound states of 
$\omega$--mesons~\cite{Tsushima,Yamazaki}
can be estimated reliably in terms of  the effective 
${\omega}N$ scattering length, which again is dominantly given by 
the $\pi^-p{\to}\omega{n}$ data near the reaction 
threshold~\cite{Klingl1}. Thus the data given in Ref.~\cite{Karami} 
directly lead to an estimate of the possible existence of $\omega$--mesonic
nuclei.

In addition, it
 should be stressed that the near-threshold production of the $\omega$--mesons in the
$pd{\to}\omega {}^3He$ reaction~\cite{Wurzinger} indicates that the
squared matrix element $|{\cal M}|^2$ strongly decreases starting from 
the center-of-mass $\omega$--meson momentum $q_\omega{\simeq}$120~MeV/c 
 to the reaction threshold where $q_\omega{=}$0. Within the 
indicated  $q_\omega$ range the reduction of $|{\cal M}|^2$ accounts
for a factor of  almost 4. As we discussed above, exactly the 
same reduction within the
same range of the $\omega$--meson momenta $q_\omega{\le}$120~MeV/c
was claimed for the $\pi^-p{\to}\omega{n}$ measurements~\cite{Karami}.
 
%The interpretation  of both observations~\cite{Karami,Wurzinger} 
%might be given in terms of a final state interaction between the 
%$\pi$--mesons from the $\omega$ decay and the recoil nucleon or
%$^3He$. The data on screening of the $\omega$--meson formation in the
%presence of the baryon field are eventually important for understanding
%of the dynamics of the production of unstable particles at near-threshold
%energies. Similar effect is well known in QCD as a color screening
%of a charmonium states in the gluon field~\cite{Matsui}, which might serve
%as a probe of the quark gluon plasma.

It should be stressed, however, that the squared matrix element $|{\cal M}|^2$ 
evaluated from the data on other heavy mesons as 
$\eta^\prime$~\cite{Sibirtsev1} and $\phi$~\cite{Sibirtsev2}  
production in $\pi{N}$ reactions did not show  such a strong 
near-threshold suppression. This difference might be ascribed to the fact that the 
$\eta^\prime$ and $\phi$--meson widths are significantly smaller than that
of the $\omega$.

The above motivation still does not cover all subjects in which
$\pi^-p{\to}\omega{n}$ data were used.
In a recent publication it was argued
that the experiments as described, e.g., in Ref. \cite{Karami} were
misinterpreted~\cite{Hanhart2}. The authors proposed a correction
factor that completely removed the above--mentioned strong energy dependence
from the cross section. If confirmed, this factor would answer
why all microscopic calculations performed so far \cite{hereweneedrefs} failed to
reproduce the energy dependence of the data correctly.
The main conclusion of this work will be to argue, that, although
the formalism derived in ref. \cite{Binnie} is correct, is was applied 
improperly in refs. \cite{Keyne,Karami,Wurzinger}. Since our
argument rests on circumstantial evidence only, we strongly call
for a remeasurment of the reaction $\pi N \to \omega N$ close
to the threshold.

%%The main message of Ref. \cite{Hanhart2} as well as this work
%is that for small outgoing momenta it is not an omega particle that
%is measured but more the continuum of its decay particles. Consequently,
%at sufficiently low outgoing momenta, the phase space can by no means
%be described by the two--body phase space, but should be given by the
% phase space of the decay particles. It should be clear that at sufficiently
%high energies the assumption of producing an omega indeed holds. In a subsequent
%section we will make these arguments more explicit.

The paper is organized as follows. In Sec.~\ref{sec:data} we summarize the 
current status of the $\pi{N}{\to}\omega{N}$ data and provide the data 
interpretation as it was given in the Ref.~\cite{Karami}. In Sec.~\ref{sec:dep}
we derive an analytical formula for the quantity measured in 
Refs.~\cite{Binnie,Keyne,Karami}. In Sec.~\ref{sec:mc}
we describe the Monte Carlo simulations of the ${\pi^-}p{\to}{\omega}n$ 
measurements. The evaluation of the imaginary part of the ${\omega}N$
scattering length from the ${\pi}N{\to}{\omega}N$ data is given in Sec.~\ref{sec:sl}.
The paper concludes with a summary.

\section{Experimental data}
\label{sec:data}
In this section we summarize the current status of the data available for the
${\pi}N{\to}{\omega}N$ total reaction cross section at pion beam energies
close to the nominal production threshold, which is given by the pole mass 
$M_\omega{=}781.94$~MeV of 
the $\omega$--meson. There are three publications from the same group that
report results for ${\pi^-}p{\to}{\omega}n$ 
measurements close to the nominal $\omega$ production 
threshold~\cite{Binnie,Keyne,Karami}. 
The data of highest quality are presented in the paper of Karami et al.~\cite{Karami},
as shown in Table~\ref{tab2}. The experimental procedure can be
sketched as follows.

The measurements were done at 33 different pion beam momenta  
$p_\pi$ scanning the range from 1040 to 1265 MeV/c.
The beam momentum resolution was $\Delta{p_\pi}/p_\pi{\simeq}$0.8\%,
so each pion momentum setting additionally covered a range 
$\simeq$8-10~MeV. At  each   pion beam 
momentum the  neutrons were collected in 60 counters set
within the angular range from 2.5$^o$ to 25.1$^o$ from the beam
direction. 

The final statistics were regrouped in 2 MeV/c wide
incident pion momentum bins with intensity about 10$^8$ pions
per bin. At each beam  momentum the final neutron momentum
$q_n$ and the neutron emission angle  $\theta^\ast$ in the 
center of mass system, as well as the flux, were determined.
The incident flux at all momenta was well fixed by weighting each event 
proportionally  to the total number of the incident pions at its particular 
momentum.  

The data were  distributed over
10 intervals of the neutron momenta $q_n$ and 10 intervals
of  $cos\theta^\ast$. This procedure is equivalent to an integration
over incoming $\pi$--meson momenta for a fixed produced neutron momentum
and angle.

For every bin in $q_n$ and $cos\theta^\ast$ the missing mass spectrum 
$d\sigma_{exp}/dm$ was reconstructed in order to separate the $\omega$--meson
spectral distribution and the background \cite{labelex}.
The $d\sigma_{exp}/dm$ spectra
were then fitted by a sum of an $\omega$--meson spectral function  and a low-order 
polynomial. The fitting allowed separation of the background and $\omega$--meson signal. 

The spectral distribution $d\sigma_{exp}/dm$ at fixed intervals
of  $q_n$ and $cos\theta^\ast$ was obtained by scanning or integrating over
the pion beam momentum $p_\pi$. To replace the integration over 
$p_\pi$ by an integration of the reconstructed missing mass spectra 
$d\sigma_{exp}/dm$ over the mass $m$, the data were corrected by a
corresponding Jacobian $\partial{p_\pi}{/}\partial{m}$. The Jacobian weakly depends 
on the beam momentum. Neither in ref. \cite{Karami} nor in ref. \cite{Keyne}
the inclusion of an additional Jacobian is mentioned. We come back to this
point below.

Finally the differential $\pi^- p{\to}\omega n$ cross sections were determined
for all 10$\times$10 intervals in neutron momentum $q_n$ and $cos\theta^\ast$. 
The differential $d\sigma/d\Omega$ cross sections shown in 
Ref.~\cite{Karami} for the different intervals of the neutron momenta 
are almost isotropic. The cross sections indicated in 
Table~\ref{tab2} were obtained by summing over the differential $d\sigma/d\Omega$
cross sections for each neutron momentum interval. The $q_n$ intervals were
specified for the central value of the cms neutron momenta $P^*$ given
within the bin ${\pm}\Delta{P}/2$.

\begin{table}[h]
\caption{The $\pi^- p{\to}\omega n$ 
cross section $\sigma$  measured by Karami et al.~\protect\cite{Karami}
for different intervals of the final neutron momenta $P^*{\pm}\Delta{P}/2$   
in the center of mass system. Also are shown the reduced cross sections 
given by the ratio $\sigma$/$P^*$.}
\label{tab2}
\vspace{1.5mm}
\begin{ruledtabular}
\begin{tabular}{cccc}
$P^*$   \hspace*{4mm} &
$\Delta{P}$  \hspace*{4mm} & $\sigma_{exp}$ \hspace*{4mm} 
& $\sigma_{exp}$/$P^*$ \vspace{1mm}\\
(MeV/c) \hspace*{4mm} &
(MeV/c) \hspace*{4mm} & (${\mu}b$)\hspace*{4mm} 
& 
(${\mu}b${}/(MeV/c))\vspace{1mm}\\
\colrule
\hline
 50 & 20 & 197$\pm$18 & 3.94$\pm$0.36 \\ 
 70 & 20 & 339$\pm$26 & 4.84$\pm$0.37 \\
 90 & 20 & 577$\pm$40 & 6.41$\pm$0.44 \\
 110 & 20 & 830$\pm$50 & 7.55$\pm$0.45\\
 130 & 20 & 1118$\pm$71 & 8.60$\pm$0.55 \\
 150 & 20 & 1350$\pm$80 & 9.00$\pm$0.53  \\
 170 & 20 & 1510$\pm$74 & 8.88$\pm$0.44 \\
 190 & 20 & 1560$\pm$83 & 8.21$\pm$0.44  \\
\end{tabular}
\end{ruledtabular}
\end{table}

The data collected in that way were interpreted as two--body  $\pi^-p{\to}{\omega}n$ 
reaction cross sections $\sigma_{2b}$. These cross sections have been 
considered as those for a stable $\omega$--meson production, since the beam 
momentum integration eliminates the dependence on the $\omega$ width. 
The two--body reaction cross section is given 
explicitly by~\cite{Byckling}
\begin{equation}
\sigma_{2b} = 6\, \frac{2}{3}\, \frac{1}{16\pi s}\,
\frac{q_n}{q_\pi} \, |{\cal M}|^2 \ ,
\label{2body}
\end{equation} 
where the factor of 6 accounts for the summation over  the number of final
spin states, the factor 2/3 is the isospin factor relating particle basis 
and isospin basis, ${\cal M}$ is the spin averaged matrix element in the isospin
basis and $s$ stands for the squared invariant collision energy.
Furthermore, in Eq.~(\ref{2body}) $q_\pi$ and $q_n$ are the incident
and final center of mass momenta, respectively, with
\begin{eqnarray}
q_\pi=\frac{\lambda^{1/2}(s,m_\pi^2,m_N^2)}{2\sqrt{s}},
\nonumber \\
q_n=\frac{\lambda^{1/2}(s,M_\omega^2,m_n^2)}{2\sqrt{s}},
\label{momenta}
\end{eqnarray}
where $m_n$ and $m_\pi$ are the nucleon and pion masses, respectively,
$M_\omega$=781.94~MeV is the pole mass of the $\omega$--meson
and the function $\lambda$ is given by
\begin{equation}
\lambda(x,y,z)=(x-y-z)^2-4yz \ .
\end{equation}

In line with Eq.~(\ref{2body}), in order  to extract the energy 
dependence of the matrix element squared $|{\cal M}|^2$ one has to divide the 
measured cross sections by the phase space volume of the two--body final state.
Already in the original publication the results on the reduced 
cross section $\sigma /P^*$ (the last column of the Table~\ref{tab2}) was 
given and the strong momentum dependence of the $\pi^{+}n{\to}\omega p$ reduced
cross section was  interpreted as a strong momentum dependence of 
the reaction matrix element.

The interpretation of the results given in Refs.~\cite{Binnie,Keyne,Karami} as a two--body
 reaction cross section was also used  in the Landoldt--B\"ornstein 
compilation of elementary reactions cross sections~\cite{LB} for both
 $\pi^- p{\to}\omega n$ and  $\pi^{+}n{\to}\omega p$  
total cross sections. There the data are shown for  fixed beam momentum
corresponding to the center of mass momenta given in the experimental source.
%with the beam momenta  calculated
%to correspond to the outgoing center of mass momenta under assumption of a
%two--body final state with the $\omega$ mass fixed at its nominal value.
% Table~\ref{k1} shows only the part of the data from the 
%Compilation~\cite{LB} for $\pi^- p{\to}\omega n$   at  laboratory 
%beam momenta $p_\pi$ from 1.097 to 1.17 GeV/c. 
Note that the final center
 of mass momenta $q_n$  given  in the experimental 
paper (here reproduced  in Table~\ref{tab2}) do not match the initial pion beam momenta
given in the compilation~\cite{LB} (here reproduced in Table~\ref{k1}) when calculated 
under the assumption, that a stable omega with its nominal mass is produced.
How the values for $p_\pi$ in Ref. \cite{LB}
where determined is unclear to us.

\begin{table}[h]
\caption{The $\pi^- p{\to}\omega n$ reaction cross section $\sigma$ and its 
uncertainty $\Delta\sigma$ as given in the Landolt-B\"ornstein 
compilation~\protect\cite{LB} at the laboratory pion beam momenta $p_\pi$ 
range from 1097 to 1170 MeV/c. Also shown are the matrix element 
squared $|{\cal M}|^2$ evaluated by Eq.~(\protect\ref{2body}) and the final neutron
momentum $q_n$ in the center of mass system given by 
Eq.~(\protect\ref{momenta}).}
\label{k1}
\vspace{1.5mm}
\begin{ruledtabular}
\begin{tabular}{ccccc}
$p_\pi$  & cross section  & error  & $|{\cal M}|^2$ & $q_n$ \\
(GeV/c) & $\sigma$ (${\mu}b$) &  $\Delta\sigma$ (${\mu}b$) 
& & (MeV/c) \vspace{1mm} \\ 
\colrule
\hline
1.097 & 197  & 18 & 182$\pm$17 & 62 \\ 
1.102 & 339  & 26 & 250$\pm$19 & 78 \\
1.109 & 577  & 40 & 347$\pm$24 & 97 \\
1.118 & 830  & 50  & 420$\pm$25 & 116 \\
1.128 & 1118 & 71  & 495$\pm$31 & 135 \\
1.140 & 1350 & 80  & 531$\pm$31 & 154 \\
1.154 & 1510 & 74 & 535$\pm$26 & 174 \\
1.170 & 1560 & 83 & 504$\pm$27 & 193  \\
\end{tabular}
\end{ruledtabular}
\end{table}
 
%Furthermore, these $\pi^- p{\to}\omega n$  data are quoted in the 
%compilation~\cite{LB} as a total reaction cross sections and are cited as 
%measured by Karami et al.~\cite{Karami}.

\begin{figure}[h]
\vspace*{-5mm}
\hspace*{-1mm}\psfig{file=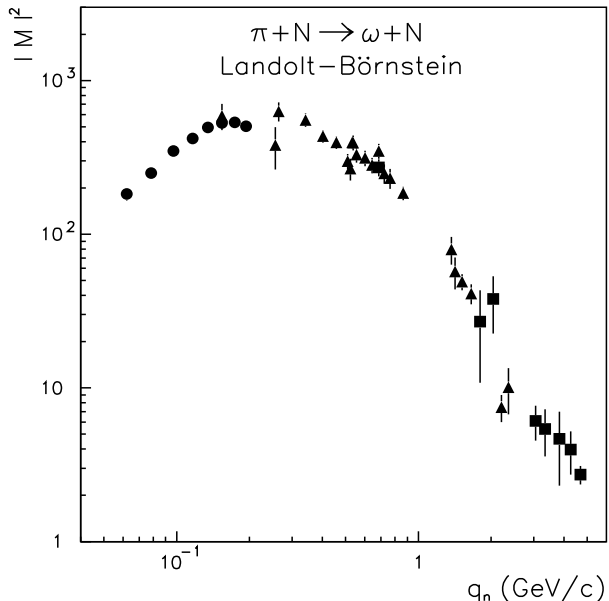,width=9.2cm,height=9cm}\vspace*{-5mm}
\caption{The spin averaged matrix element squared $|{\cal M}|^2$ evaluated 
by Eq.~(\protect\ref{2body}) from the total $\pi^- p{\to}\omega n$ (triangles) and  
$\pi^{+}n{\to}\omega p$ (squares) cross sections taken from the 
compilation~\cite{LB}  as a function of the final neutron
momentum $q_n$ in the center of mass system given by 
Eq.~(\protect\ref{momenta}). The $\pi^- p{\to}\omega n$ data quoted~\cite{LB} for 
the laboratory pion beam momenta $1.097{\le}p_\pi{\le}1.17$ GeV/c
are shown by the circles.}
\label{figmatele}
\end{figure}

The spin averaged squared matrix element  $|{\cal M}|^2$ defined through 
Eq.~(\ref{2body}) is shown in Fig.\ref{figmatele} as well
as in Table \ref{k1} as a function of the neutron 
momentum $q_n$. The results are shown for the $\pi^- p{\to}\omega n$ (triangles) and  
$\pi^{+}n{\to}\omega p$ (squares) reactions. The data~\cite{LB} on the $\pi^- p{\to}\omega n$ 
reaction for laboratory pion beam momenta between 1.097 and 1.17~GeV/c are shown
by the circles.  

Fig.\ref{figmatele} illustrates the very strong momentum dependence of 
the squared matrix element  near the reaction threshold 
at $q_n{\le}$200~MeV/c mentioned above.
 Within the short range of the neutron momentum $q_n$ from
154 to 62~MeV/c $|{\cal M}|^2$ decreases almost by a factor of 3.

Summarizing, there are three potential sources of energy dependence of 
the $\pi^- p{\to}\omega n$ total cross section close to the nominal $\omega$ 
production threshold: the phase space, the width of the $\omega$--meson and
the  $\pi^- p{\to}\omega n$ transition amplitude. It is the last quantity 
$|{\cal M}|^2$ that contains all the $\omega$ production dynamics. 
In the energy regime of interest here the width effect on the energy
dependence of the cross section is quite large. As was described above, to 
remove this effect from the data in a model--independent way, the 
authors of Ref.\cite{Binnie,Keyne,Karami} proposed to integrate over some range of 
initial pion momenta $p_\pi$ for a fixed final neutron momentum labeled $P^*$.
It was shown in ref. \cite{Binnie} how the count rates can be related to 
the cross sections.
Due to the aforementioned difficulties for model studies to 
account for the resulting energy dependence of the cross section we
repeat here the derivation from Binnie et al. and check the approximations
made along the line with an event--by--event simulation.

\section{Energy dependence of the cross section}
\label{sec:dep}
In Ref.~\cite{Hanhart2} it was argued that the energy dependence of the
$\pi^- p{\to}\omega n$ reaction matrix element as shown in 
Fig. \ref{figmatele} can be traced to a misleading 
interpretation of the measured count rates. The basic findings
were that it is indeed  possible to extract the transition matrix element independent 
of the shape of the mass distribution of the resonance, however---in
 contradiction to the claims of Refs.\cite{Binnie,Keyne,Karami}---it was
argued that the energy dependence of the count rates is different 
from that of the two--body reaction. As a consequence it was claimed,
that the above mentioned energy dependence of the matrix element squared
$|{\cal M}|^2$ is a purely kinematical effect. In this section we
want to check this claim by repeating the derivation of Binnie et al..
As it will turn out, the theoretical part of Ref. \cite{Binnie} is
correct, however, there is a chance of misinterpretation of the
final formula (formula (10) in Ref. \cite{Binnie}) which might have
lead to a misuse of the technique. We will come back to this point below.
In addition, we regard the following discussion as usefull for
future experiments for it is somewhat more compact than the original
presentation of ref. \cite{Binnie}.

Let us start with the cross section for ${\pi}N{\to}XN$ reaction, where
$X$ denotes the decay products of the unstable meson. The reaction cross 
section can be expressed as
\begin{equation}
d\sigma = (2\pi)^4 \, \frac{4}{q_\pi\sqrt{s}} \,
|{\cal M}|^2 d\zeta\ .
\label{dsig}
\end{equation}
As in Eq.~(\ref{2body})  $q_\pi$ denotes the cms momentum of the initial 
pion and ${\cal M}$ the spin averaged invariant matrix element. The factor of 
four is a combination of spin and isospin factors, as is given by Eq.~(\ref{2body}).
The trivial energy dependences of the reaction cross section 
are collected in the function $\zeta$, which is defined as
\begin{equation}
d\zeta = d\Phi_{k+1}(p;p_n,p_1,...,p_k) \,\, |D(m^2)W(p_1,...,p_k)|^2 \ ,
\label{dzeta}
\end{equation} 
where we assume the unstable meson, whose propagation is described by $D(m^2)$,
to decay into the $k$ particles through the vertex function $W$. Here $m^2$ is the total
invariant mass of the $k$ final decay particles with $m^2=(\sum p_i)^2$,
where the $p_i$, $i=1..k$ denote the corresponding four--momenta.
Furthermore, the total initial four-momentum is denoted by $p$. The phase space 
of the final $k$ particles and the final nucleon with the four-momentum
$p_n$ is defined as 
\begin{eqnarray}
d\Phi_{k+1}(p;p_n,p_1,...,p_k) = \delta^4(p-p_n-\sum p_i) \nonumber \\
\times \frac{d^3p_n}{(2\pi)^3}\frac{1}{2E_n}
\prod_i \frac{d^3p_i}{(2\pi)^3 2\omega_i},
\end{eqnarray}
where $E_n$ and $\omega_i$ are the energy of the final nucleon and $i$-th
decay particle, respectively.

Using the unitarity relation we can introduce the spectral function $\rho$ as
\begin{eqnarray}
\nonumber
(2\pi )^3\int d\Phi_k(p_X;p_1,...,p_k)\, \, |D(m^2)W|^2 = \\
 -\frac{1}{\pi}\Im D(m^2) =: \rho(m^2).
\end{eqnarray}
A standard choice for the spectral function is that of a
Breit--Wigner resonance,
\begin{equation}
\rho (m^2) = \frac{1}{\pi} \frac{M_\omega\Gamma_\omega}
{(m^2-M_\omega^2)^2+M_\omega^2\Gamma_\omega^2} \ ,
\label{BW}
\end{equation}
where $M_\omega$ and $\Gamma_\omega$ are the pole mass and width
of the $\omega$--meson, respectively. It is this form that we
will use in the evaluation of intermediate results. For
the final result, however,
the exact shape of the spectral function  is irrelevant.
All that we will use
is the normalization 
condition, namely
\begin{equation}
\int dm^2 \, \rho (m^2) = 1.
\label{norm}
\end{equation}
Now we can rewrite Eq.~(\ref{dzeta}) in the center of mass system as
\begin{eqnarray}
\nonumber
d\zeta &=& \frac{1}{4(2\pi)^6}\, \rho(m^2) \, dm^2 
\frac{d^3q_n}{\omega  E_n}
\delta(\sqrt{s}-\omega -E_n) \\
&=& \frac{1}{2(2\pi)^6}\, \rho(s-2\sqrt{s}E_n+m_n^2)\, \frac{d^3{q_n} }{E_n},
\label{dzetaresQ}
\end{eqnarray}
where $\omega{=}\sqrt{m^2+\vec {q_n}{}^2}$ is the energy of the final 
unstable meson state with the total invariant mass $m$ and $q_n$ denotes
the momentum of the final nucleon in the center of mass system.
%Note that the argument of the spectral function is given by
%$s-2\sqrt{s}E_n+m_n^2$, while $\rho$ is peaked around the physical pole mass 
%of the unstable final state. 

Finally, by introducing Eq.~(\ref{dzetaresQ}) into the  Eq.~(\ref{dsig}), we 
obtain  the differential reaction cross section as a function of the
nucleon momentum:
\begin{equation}
\frac{d\sigma}{dq_n} =
\frac{1}{2\pi q_\pi \sqrt{s}}\, \frac{q_n^2}{E_n}\, \, |{\cal M}|^2 \,\,
\rho(s-2\sqrt{s}E_{q_n}+m_n^2).
\label{dsigfinal}
\end{equation}

This formula agrees with that used in Ref.~\cite{Sibirtsev2} up to differences in
the normalisation. 
Note that in the limit of vanishing width of the unstable particle
Eq.~(\ref{dsigfinal}) transforms into the two--body cross section of Eq.~(\ref{2body}) 
for the production of particles with fixed masses, since
\begin{equation}
\lim_{\Gamma{\to}0} \rho(s-2\sqrt{s}E_n+m_n^2) = \frac{E_n^\prime}
{2\sqrt{s}}\left(\frac{1}{q_n^\prime}\right) \, \delta (q_n-q_n^\prime),
\label{gamlim}
\end{equation}
where $E_n^\prime$ and $q_n^\prime$ denote the energy and relative momentum of 
the produced stable particles.
It is interesting to investigate for what values of $\Gamma$ the right
hand side of Eq. (\ref{gamlim}) is a good approximation to the
left hand side. Naturally, the parameter that controls the behavior
of the cross section should depend on the energy and the width of the
decaying particle as well as the resolution of the detector.
For the cross section to appear as a two--body cross section we should not 
resolve the decay particles any further. In Ref. \cite{Hanhart2} it was shown
that for 
\begin{equation}
\frac{2 P^* \Delta P}{\mu \Gamma} \ \gg 1 \ ,
\label{xi}
\end{equation}
where $\mu$ denotes the reduced mass of the $\omega n$ system,
the reaction rate behaves indeed like a two--body cross section.
It should be obvious that the ratio $\Delta P/\Gamma$ appears here,
 for this ratio measures how closely we look at the production rates.
 For a given detector
resolution $\Delta P$ we thus deduce a critical value for $P^*$ below which 
deviations from the two--body behaviour should be expected on purely kinematical
grounds. Using $\Delta P = 20$ MeV as given in Table \ref{tab2} we find that, for 
values of $P^* \gg 90$ MeV, the cross section for a fixed energy should behave
like a two--body cross section.

\begin{figure}[t]
\vspace*{-6mm}\hspace*{-3mm}\psfig{file=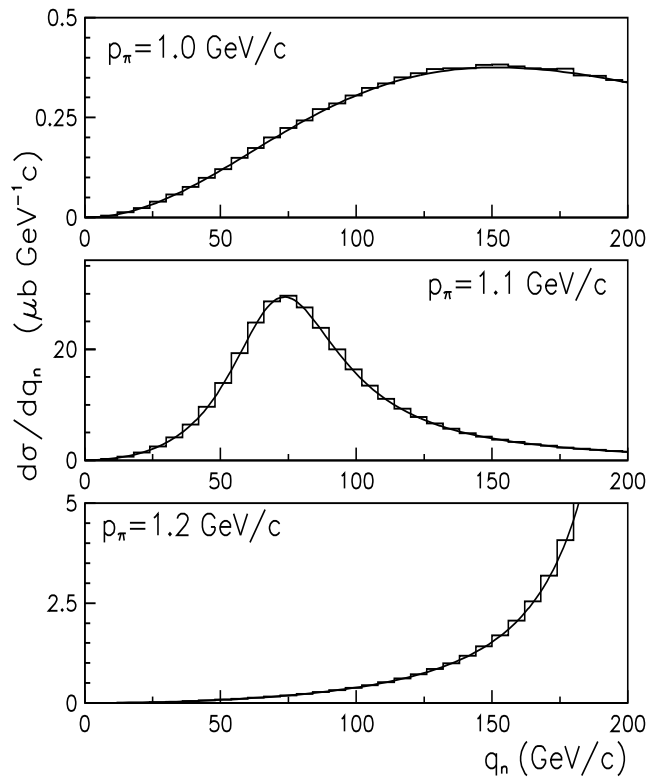,width=9.5cm,height=11.4cm}\vspace*{-5mm}
\caption{The neutron momentum spectra calculated at different
fixed pion beam momenta $p_\pi$. The histograms show the results
from the event-by-even simulations, while the solid lines indicate
the Eq.~(\protect\ref{momenta}). The spectra are shown only for the 
neutron momenta  $q_n$ range covered  by Karami 
experiment~\protect\cite{Karami}.}
\label{dsigoqnsfixed}
\end{figure}

The solid lines in the Fig.~\ref{dsigoqnsfixed} show the differential 
cross section $d\sigma/dq_n$ calculated for the $\pi^- p{\to}\omega n$ 
reaction at fixed $\pi$--meson momenta $p_\pi$. Here we employed the 
relativistic Breit--Wigner form as given in Eq.~(\ref{BW})
  for the $\omega$--meson spectral function.
 Furthermore, in this calculation, we 
employed the almost constant matrix element squared $|{\cal M}|^2$, 
as defined in Eq.~(\ref{matrix1}). The neutron spectra are shown in 
 Fig.~\ref{dsigoqnsfixed} for the $q_n$ range studied by 
Karami et al.~\cite{Karami}.
Fig.~\ref{dsigoqnsfixed} shows that the differential cross section
$d\sigma/dq_n$  reflects the 
behavior of the spectral function. 
%Furthermore, it is clear that
%the pion beam with a highest momentum produces mostly energetic neutrons,
%while the low energy pions produce the neutrons with the momenta 
%distributed over the full $q_n$ range measured in Ref.~\cite{Karami}.  

Fig.~\ref{dsigoqnsfixed} illustrates an important feature: namely,
the differential cross section at different pion momenta, but at the same fixed
neutron momentum are substantially different in absolute value. 
Naively one would expect
 that the $d\sigma/dq_n$ are identical for different
$p_\pi$ but  the same $q_n$. Note that the difference in the flux factor
due to the different pion momenta is almost negligible within the range
$1{\le}p_\pi{\le}1.2$~GeV/c. The calculations by Eq.~(\ref{momenta}) clearly show
an absolutely  different situation, since the phase space dependence of the
cross section can not be factored from the spectral function
density. 

%Obviously, the contribution to the $\pi^- p{\to}\omega n$ cross section
%at fixed value of the neutron momentum $q_n$ from the different $\pi$--meson
%momenta is not only inversely proportional to the flux factor (which is
%linear in the final cms momentum labeled as $P^*$), as was
%assumed in Refs.~\cite{Binnie,Keyne,Karami}. 

It was the idea of Refs.\cite{Binnie,Keyne,Karami} to remove the 
dependence on the spectral function from Eq.~(\ref{dsigfinal}) through  an 
integration over the initial pion momentum $p_\pi$. The necessary assumption 
 is that the flux as well as the matrix element 
squared $|{\cal M}|^2$ only weakly depend on the total collision energy
when evaluated within a limited range of final $q_n$ momenta.
In addition, we need to assume that neither of the two vary significantly
when $m$ is varied within the range where the spectral function is
large. 
%The experimental range of momenta was $\Delta P = 20$ MeV \cite{Karami}. 
%The matrix element we extract varies on a much larger momentum scale (c.f. Eq.
%(\ref{matrix1})), showing the consistency of the approach.

By integrating Eq.~(\ref{dsigoqnsfixed}) over the laboratory $\pi$--meson
momentum $p_\pi$ we obtain  the $\pi^- p{\to}\omega n$ cross section
for the range of the neutron cms momentum $P^*{\pm}\Delta{P}/2$ as
\begin{equation}
{\bar \sigma (P^*,\Delta{P})} =
\int_{p_\pi^-}^{p_\pi^+}\!\!\!\!dp_\pi \,\, \frac{1}{\Delta P}
\int_{q_-}^{q_+}\!\!\!\!dq_n
\, \frac{d\sigma}{dq_n},
\label{dsigeint1}
\end{equation}
where the limits of the integration over the neutron momentum are fixed by
\begin{equation}
q_{\pm}=P^*{\pm}\Delta{P}/2, 
\end{equation}
while the integration over the pion laboratory momentum were
performed~\cite{Karami} within the range  from  1040 to 1265~MeV/c.

\begin{figure}[h]
\vspace*{-6mm}\hspace*{-1mm}\psfig{file=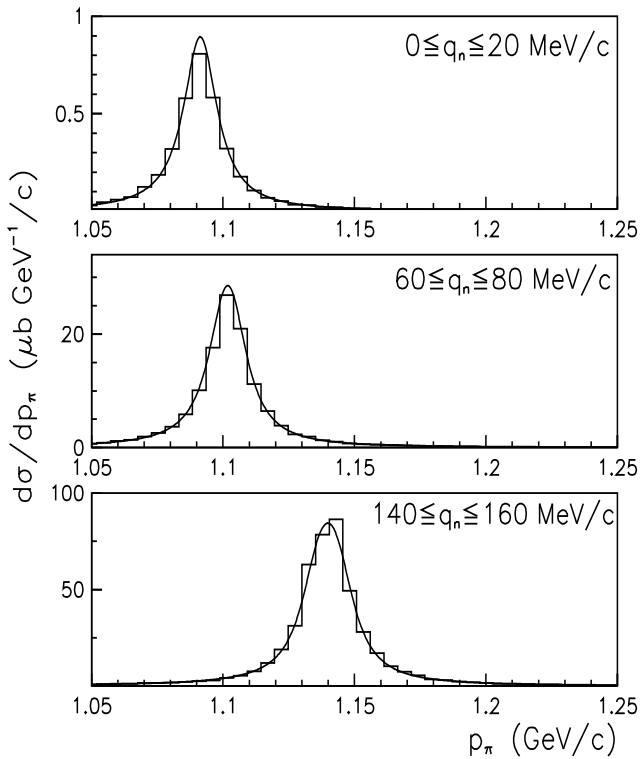,width=9.2cm,height=11cm}\vspace*{-5mm}
\caption{The  differential cross section as a function of 
the pion beam momentum $p_\pi$ calculated for  different ranges of  neutron 
momenta $q_n$. The histograms are results from the event-by-event generator
while the solid lines show the calculations using Eq.~(\protect\ref{dsigeint1}).}
\label{dsodp}
\end{figure}

The solid lines in  Fig.~\ref{dsodp} show the neutron differential 
cross section as a function of the $\pi$--meson beam momentum calculated
for the two different intervals of the neutron momenta. The shapes of the
$d\sigma/dp_\pi$ distributions reflects the Breit--Wigner spectral function
distribution. The width is dominantly given by the $\omega$--width with
a tiny correction from the spread in the outgoing momentum.

%Furthermore, it is clear from the Fig.~\ref{dsodp} that 
%the $p_\pi$ integration over the range specified by an experimental conditions 
%might be replaced by an integration over an infinite range.  

To remove  the dependence on the spectral function $\rho$, we replace 
in Eq.~(\ref{dsigeint1}) the integration over $p_\pi$ by an integration over 
the  invariant mass squared $m^2$ and additionally impose the normalization
condition of
Eq.~(\ref{norm}). As illustrated in Fig.~\ref{dsodp} the $p_\pi$ integration 
can be assumed to be done  over an infinite range. Therefore,
 the $\pi^- p{\to}\omega n$ 
cross section is given as
\begin{equation}
{\bar \sigma (P^*,\Delta{P})} =
\kappa
\left[ P^*{}^2+\frac{1}{12}(\Delta P)^2\right] 
|{\cal M}|^2 \ ,
\label{dsigeint}
\end{equation}
with 
\begin{equation}
\kappa = \frac{1}{2\pi E_n q_\pi \sqrt{s}}
\, \, \left(\frac{\partial p_\pi}{\partial m^2}\right),
\end{equation}
where $E_n{=}\sqrt{{P^*}^2+m_n^2}$ and
$s$ denotes the invariant collision energy evaluated for the production of 
a stable $\omega$--meson  with center of mass momentum $P^*$.
The weakly energy dependent Jacobian is given by
\begin{equation}
\left(\frac{\partial p_\pi}{\partial m^2}\right) \simeq \frac{s-m_\pi^2-m_n^2}{4M_\omega 
m_nq_\pi},
\label{jac}
\end{equation}
where we have used the non relativistic expression for the total energy of the 
$\omega$--meson.

The expression given in formula (\ref{dsigeint}) agrees to eq. (10) of ref.
 \cite{Binnie} for an ideal detector and beam, 
which states, translated into the quantities used above
\begin{equation}
\bar \sigma =  2M_\omega\left(\frac{\partial p_\pi}{\partial m^2}\right)
\frac{1}{\Delta P}\int_{\Delta \tau} d\tau dx^L J_\tau \frac{d\sigma}{dt}
 \ ,
\label{sbbin}
\end{equation}
where $x^L$ denotes the cosine of the scattering angle in the lab frame,
$\tau$ denotes the time of flight of the outgoing neutron and $J_\tau$ is the
Jacobian relating the system $(\cos (\theta ),\tau )$ to $(m,t)$ (c.f. eq. (6) in ref.
\cite{Binnie})
\begin{equation}
J_\tau = \frac{2(q_n^L)^4p_\pi m_N}{(E_n^L)^2M_\omega d} \ ,
\label{jtau}
\end{equation}
where $d$ denotes the flight distance for the neutrons and $q_n^L(E_n^L)$ denote
the neutron momentum (energy) in the laboratory frame. In addition,
for an isotropic matrixelement
\begin{eqnarray}
\nonumber
\frac{d\sigma}{dt} &=& \left(\frac{\pi E_n}{2q_\pi q_n^2\sqrt{s}}\right)\int dm^2
 \frac{d^3\sigma}{dq_nd\Omega} \\
&=& \frac{1}{16\pi q_\pi^2 s}|{\cal M}|^2 \ .
\end{eqnarray}
Thus, at this stage it seems as if the experiments of refs. 
\cite{Binnie,Keyne,Karami,Wurzinger} where analysed properly.
However, from what is written in refs. \cite{Keyne,Karami} it is unclear,
if $\sigma_{exp}$, defined through
\begin{equation}
\bar \sigma (P^*,\Delta{P}) =2M_\omega
\,\left(\frac{\partial p_\pi}{\partial m^2}\right)\,  \sigma_{exp}(P^*).
\label{sigomdef}
\end{equation}
is given, or $d\sigma /dt$ as it can be extracted using eq. (\ref{sbbin}).
The former option corresponds to interpreting $J_\tau$
in eq. (\ref{sbbin})
 as $J_{(\tau,q_n)}$---the Jacobian that connects $(\cos (\theta ),\tau )$ 
to $(q_n,t)$, where
\begin{equation}
J_{(\tau,q_n)} = \frac{\tilde \mu}{q_n}J_{\tau} \ ,
\label{jtauq}
\end{equation}
where $\tilde \mu = M_\omega E_n/\sqrt{s}$, which reduces to the reduced mass
of the final state in the non--relativistic limit.
This interpretation of the Jacobian appearing in eq. (\ref{sbbin}) is natural
(though wrong) to the extend that the cut in the outgoing neutron momentum
is given in terms of $q_n$, whereas eq. (\ref{sbbin}) is formulated in
terms of the time of flight $\tau$.
The following evidence supports, that this misinterpretation acctually
took place, namely
\begin{itemize}
\item both publications \cite{Keyne,Karami}
 only mention that there is a Jacobian to be
included without further comment wether they used
$J_\tau$ (given in eq. (\ref{jtau}) or $J_{(\tau,q_n)}$ (given in eq. (\ref{jtauq}),
 thus suggesting that only the naive one that
connects the coordinate systems---namely $J_ {(\tau,q_n)}$ is used and
\item in ref. \cite{Wurzinger} there is explicitly only the analog of $J_{(\tau,q_n)}$
mentioned and no other. Note, the momentum dependence of the amplitude given
in ref. \cite{Wurzinger} is consistent with that by Karami et al. \cite{Uzikov}.
\end{itemize}
Thus one may consider two options:
{\it Either} $J_\tau$ (c.f. eq. (\ref{jtau}))
 was indeed included in the experimental analysis
of Refs. \cite{Binnie,Keyne,Karami,Wurzinger} and the data. Then
the strong energy dependence as shown in Fig. \ref{figmatele} is physical.

{\it Or} the experimental data were analysed using $J_{(\tau,q_n)}$ as given
by eq. (\ref{jtauq}).

In what follows we investigate the consequences 
of the second option. In other words, we study the implications
of the {\it assumption}
that it was $\sigma_{exp}$ defined in eq. (\ref{sigomdef}) 
that was given in the experimental papers
\cite{Keyne,Karami}.

Matching Eqs.~(\ref{sigomdef})  and (\ref{dsigeint}), we arrive at the 
central formula of this manuscript, providing the relation between the 
measured cross section~\cite{Binnie,Keyne,Karami} and the squared matrix element 
$|{\cal M}|^2$ of the $\pi^-p{\to}\omega{n}$ reaction, namely
\begin{equation}
\sigma_{exp} = \frac{1}{4\pi q_\pi s \tilde \mu}
\,\, \left({P^*{}^2+(\Delta P)^2/12 }\right)
\,\,\, |{\cal M}|^2 \ .
\label{elma}
\end{equation}
One might wonder why the resolution of the neutron detector enters
the final expression. On a second thought this should be quite 
obvious, since for every given value of $\Delta P$ there will be a count rate even
for a minimal value of $P^*$. This is most clearly demonstrated in 
the uppermost panel of Fig. (\ref{dsodp}).

By construction,  ${\cal M}$  is the matrix element
for the reaction $\pi^-p{\to}\omega{n}$ considering
the $\omega$--meson  as a stable particle, since the width
of the omega was treated explicitly.
We can therefore define an  effective two--body
cross section $\sigma^{eff}$ given by Eq.~(\ref{2body}) in terms
of the cross section $\sigma_{exp}$ measured in Ref.~\cite{Karami}: 
\begin{equation}
\sigma^{eff} = 
\frac{\tilde \mu P^*}{{P^*}^2+(\Delta P)^2/12}
\,\,\, \sigma_{exp} \ ,
\label{effsig}
\end{equation}

Therefore, the effective cross section $\sigma^{eff}$ is the $\pi^-p{\to}\omega{n}$
total reaction cross section taken under the assumption that the produced
$\omega$--meson is stable with the fixed mass $M_\omega{=}$781.94~MeV. 
In Table~\ref{tabsigeff} we provide the corrected cross section $\sigma^{eff}$
based on the hypothesis that the formalism given in Ref. \cite{Binnie}
was applied improperly.
The corrected $\pi^-p{\to}\omega{n}$ reaction cross sections
deviate from the results quoted in the compilation~\cite{LB} by a factor
up to 8 within the range of pion beam momenta 
$1094{\le}p_\pi{\le}1167$~MeV/c.

\begin{table}[h]
\caption{The corrected effective $\pi^-p{\to}\omega{n}$ 
cross section $\sigma^{eff}$ as derived by us. The second to last column contains
the error as it is derived directly from the data~\cite{Karami}, 
whereas in the error given in the last column the uncertainty 
of the neutron momentum $P^*$ is included as well. The relation between the central
neutron momentum $P^*$ and the pion beam momentum $p_\pi$ is given
by Eq.~(\ref{momenta}).}
\label{tabsigeff}
\vspace{1.5mm}
\begin{ruledtabular}
\begin{tabular}{lllll}
{\hspace*{2mm}$P^*$ \hspace*{2mm}} & {$p_\pi$ \hspace*{4mm}} 
& {$\sigma^{eff}$ \hspace*{4mm}} 
& $\delta\sigma^{eff}_{exp}$ \hspace*{6mm} &
$\delta \sigma^{eff}_{exp+P^*}$ \\
 (MeV/c) & (MeV/c) & (${\mu}b$) & (${\mu}b$) & (${\mu}b$)\\
\colrule
       50 &   1094 &   1656 &    151 &    364 \\
       70 &   1099 &   2049 &    157 &    332 \\
       90 &   1106 &   2720 &    189 &    356 \\
      110 &   1115 &   3206 &    193 &    350 \\
      130 &   1125 &   3656 &    232 &    365 \\
      150 &   1137 &   3828 &    227 &    341 \\
      170 &   1151 &   3780 &    185 &    289 \\
      190 &   1167 &   3494 &    186 &    262
\end{tabular}
\end{ruledtabular}
\end{table}

\begin{figure}[t]
%\vspace*{5mm}
\hspace*{-2mm}\psfig{file=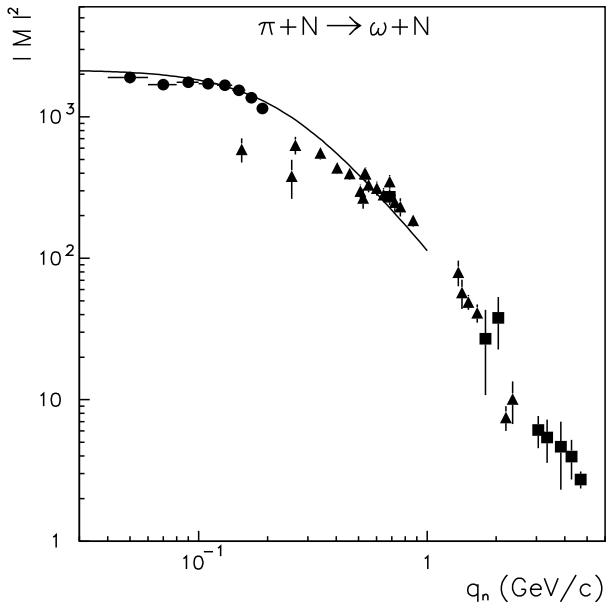,width=9.2cm,height=9cm}\vspace{-5mm}
\caption{The  spin averaged matrix element squared $|{\cal M}|^2$ 
extracted  from the total $\pi^- p{\to}\omega n$ (triangles) and  
$\pi^{+}n{\to}\omega p$ (squares) cross sections taken from the 
compilation~\cite{LB}  as a function of the final neutron
momentum $q_n$. The circles show the corrected matrix element
squared evaluated using Eq.~(\protect\ref{elma}) for the data~\protect\cite{Karami} 
on the $\pi^- p{\to}\omega n$ reaction. 
The solid line shows the parametrization
of the matrix element given by Eq.~( \protect{\ref{matrix1}}) with the
parameters fixed by a fit of the event--by--event simulation of the data
of Ref.~\protect\cite{Karami}.}
\label{figwe}
\end{figure}

We can now evaluate  the corrected matrix element squared 
$|{\cal M}|^2$ from the $\pi^-p{\to}\omega{n}$ data
employing  Eq.~(\ref{elma}). 
The results are shown in Fig.~\ref{figwe} and clearly illustrate
that the corrected matrix element is in line with the other data and
does not indicate any pathological energy dependence at small neutron momenta.
The resulting energy dependence is in a lot better agreement with that of model
studies \cite{hereweneedrefs}. This could be taken as further evidence, that
indeed the formalism derived in Ref. \cite{Binnie} was applied improperly.

\section{event--by--event simulations}
\label{sec:mc}
In addition, to check the analytical calculations of the previous section,
we also developed an event--by--event generator to simulate the experimental
measurements described in Refs.~\cite{Binnie,Keyne,Karami}. In addition, this
simulation allows us to discuss several aspects of the technique developed
in Ref. \cite{Binnie} in detail.

In accordance with  Ref.~\cite{Karami}, the 
$\pi^- p{\to}\omega n$ events were simulated at pion momenta 
randomly selected within the range between 1040 and 1265
MeV/c. At fixed pion momentum $p_\pi$ the events  were generated by
the following method.

1. The flux factor was determined as
\begin{equation}
{\cal L} = 8\, \pi^2 \, \lambda^{1/2}(s,m_\pi^2,m_p^2),
\end{equation}
where $m_\pi$, $m_p$ and $m_n$ are the pion, proton and neutron masses, respectively
and $s$ is the squared invariant energy of the event given by
\begin{equation}
s=m_n^2 + m_\pi^2 + 2m_n \sqrt{p_\pi^2+m_n^2}.
\end{equation}

2. The squared $\omega$--meson mass $m^2$ was randomly generated by
the Breit-Wigner probability distribution as defined in Eq.~(\ref{BW}).
%\begin{equation}
%BW(m^2)=\frac{1}{\pi} \, \, \frac{\Gamma_\omega \,  M_\omega}
%{(m^2-M_\omega^2)^2 + \Gamma_\omega ^2 M_\omega^2},
%\end{equation}
%where $M_\omega$ and $\Gamma_\omega$ are the pole mass and width
%of the $\omega$--meson, respectively.

3. The neutron momentum was calculated as
\begin{equation}
q_n=\frac{\lambda^{1/2}(s,m^2,m_n^2)}{2\sqrt{s}},
\label{qnofm}
\end{equation}
and the phase space factor 
of the event is given as
\begin{equation}
\Phi = \pi \, \frac{\lambda^{1/2}(s,m^2,m_n^2)}{2s}.
\end{equation}

4. Now each event at fixed pion beam momentum $p_\pi$ and 
with certain $\omega$--meson mass $m$ was accounted for with
the weight~\cite{Byckling}
\begin{equation}
W_{ev}=\frac{\Phi}{\cal L}\, \, 4\, |{\cal M}|^2 \ .
\end{equation}
As in Eq.~(\ref{2body}), ${\cal M}$ denotes the spin averaged matrix element in 
isospin basis and  the factor of 4 stems from a combination of the number of 
final spin states and the isospin factor.
Note that this expression agrees with that for a two--body cross section 
(c.f. Eq.~(\ref{2body})), however, in our analysis we treat the energy and
mass dependence of $q_n$ properly.

5. In order to ensure energy conservation  we impose 
the kinematical condition $W_{ev}$=0 when $\sqrt{s}{\le}m{+}m_n$.

\begin{figure}[h]
\vspace*{-7mm}\hspace*{-1mm}\psfig{file=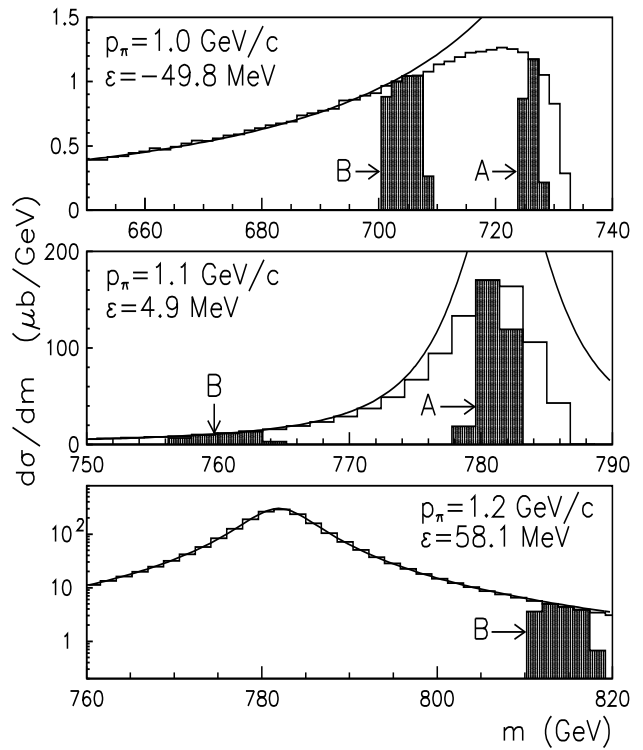,width=9.2cm,height=11cm}\vspace*{-5mm}
\caption{The missing mass spectra calculated for the fixed pion
momenta $p_\pi$. The full histograms show the results from the event-by-even 
generator. The solid lines indicate the Breit--Wigner distribution for the
$\omega$--meson. The hatched histograms show the mass spectra for the
neutron momenta ranges $60{\le}q_n{\le}80$~MeV/c (A) and 
$140{\le}q_n{\le}160$~MeV/c (B). 
The excess energy $\epsilon = \sqrt{s}-m_n-M_\omega$.}
\label{bin7}
\end{figure}

Now, to calculate the $\pi^- p{\to}\omega n$ cross section at fixed
pion momentum $\sigma(p_\pi)$, we generate $N$ events  and sum them as
\begin{equation}
\sigma(p_\pi) =\frac{1}{N} \sum_{ev=1}^{ev=N} W_{ev}.
\end{equation}

As described in the previous section, in Ref.~\cite{Karami}
the data are given not at fixed pion beam momenta, but at
different ranges of the neutron momenta. At fixed pion beam momentum 
the reaction $\pi^-p{\to}{\omega}n$ produces the neutron momentum
spectrum rather than a fixed neutron momentum because of the
variation of the $\omega$--meson mass. It can be well
understood within the event--by--event simulations when proceeding from 
step 2 to step 3, as illustrated in Fig.~\ref{dsigoqnsfixed}. The histograms
in Fig.~\ref{dsigoqnsfixed} show the results from the Monte Carlo simulations,
which are in agreement with the analytical evaluations.

Fig.~\ref{bin7} also shows the missing mass distribution simulated 
for the $\pi^-p{\to}{\omega}n$ reaction at different fixed pion beam momenta.
The histograms in Fig.~\ref{bin7} indicate the total missing mass spectra, 
while the solid lines indicate the relativistic Breit--Wigner distribution 
for the $\omega$--meson. It is clear that at low pion momenta only 
$\omega$--mesons with small masses can be produced because of the
energy conservation imposed by the condition of step 5. 

The hatched histograms in  Fig.~\ref{bin7}, moreover, 
show the missing mass spectra for the two neutron momenta 
ranges $60{\le}q_n{\le}80$~MeV/c (A) and $140{\le}q_n{\le}160$~MeV/c (B).
Therefore, for a fixed range of final neutron momenta, the 
contribution from the different $\omega$--meson masses comes due to the
different incident pion momenta.  
Fig.~\ref{bin7} most clearly illustrates the basic idea of the
measurements~\cite{Binnie,Keyne,Karami}: namely to saturate the $\omega$--meson 
spectral function by scanning the pion beam momentum. As 
indicated by the hatched histograms, the $\omega$ production cross sections
at different $\pi$--meson momenta but at the same neutron momentum $q_n$
substantially differ in absolute value.  

An additional generation over the inital pion momentum
 $p_\pi$ is required, randomly 
scanning the experimental pion momentum range.
After introducing the Jacobian defined in Eq.~(\ref{jac}), we
 fit the experimental data~\cite{Karami} in order to 
extract the matrix element squared $|{\cal M}|^2$ as well as its
energy dependence. 
For the squared matrix element $|{\cal M}|^2$ we assume an energy dependence  given by
\begin{equation}
|{\cal M}|^2 = \frac{M_0^2}{1+bq_n^2},
\label{matrix1}
\end{equation}
with parameters $M_0$ and $b$ to be fit to the  data of Ref. ~\cite{Karami}.

\begin{figure}[t]
%\vspace*{-5mm}
\hspace*{-2mm}\psfig{file=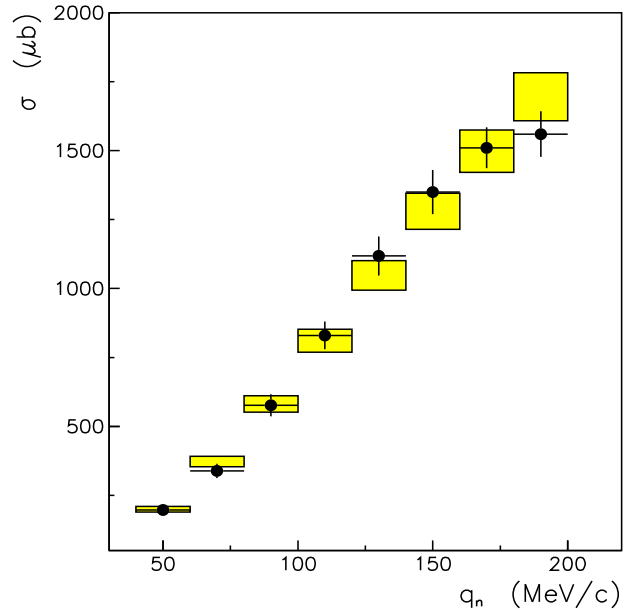,width=9.4cm,height=9cm}\vspace*{-5mm}
\caption{The cross sections $\sigma_{exp}$ measured at Karami 
experiment~\cite{Karami} at 
different intervals of a neutron momenta $q_n$. The experimental 
results~\protect\cite{Karami} are shown by solid circles, while the boxes
indicate our calculations. The size of the boxes illustrates the uncertainties
in evaluation of the reaction matrix element $|{\cal M}|$ and the neutron
momentum interval.}
\label{bin1}
\end{figure}

Fig.~\ref{bin1} shows the experimental results from Ref.~\cite{Karami}
together with the results of the fit with the parameters
\begin{equation}
M_0^2=(2.1 \pm 0.1)\times 10^3 \ \mbox{and}\ b=(0.7\pm0.1) \ \mbox{fm}^2 \ .
\label{mafit}
\end{equation}
The experimental cross sections $\sigma_{exp}$~\cite{Karami} are shown
by the solid circles, while the boxes are our calculations. The size of
the boxes shows the uncertainty in the extraction of the matrix element
squared $|{\cal M}|^2$ as well as the range of integration used for the neutron momentum
as given by $\Delta{P}$.

Finally, the squared matrix element parametrized by
Eqs. (\ref{matrix1}) and (\ref{mafit}) fitted by using the Monte Carlo simulations to the
experimental data~\cite{Karami} is shown by the solid line in 
the Fig.~\ref{figwe}. It agrees well
 with the results derived analytically. We also find that the
parametrization given by Eq.~(\ref{matrix1}) works rather well over
a wide energy range. Note that the fit was carried out on the basis
the dataset of Ref. \cite{Karami} only, thus only the momentum range from
50 to 200 MeV was fit.

\section{Evaluation of the scattering length}
\label{sec:sl}
\begin{figure}[h]
\vspace*{5mm}
\hspace*{-5mm}\psfig{file=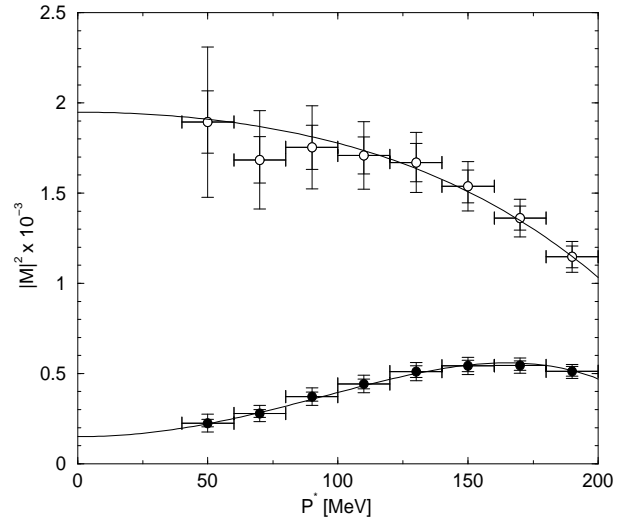,width=8cm,height=7cm}\vspace{-5mm}
\caption{The matrix element extracted from the data of Ref.~\protect{\cite{Karami}}.
The opaque circles denote the corrected matrix element
using Eq.~(\protect{\ref{effsig}}) whereas the filled circles
show the uncorrected result. The small error bars are those from
the data only, whereas the large error bars include the uncertainty in $P^*$, given
by ${\Delta}P{=}20$~MeV. The solid lines are polynomial fits to the data.}
\label{fig}
\end{figure}

The matrix element at vanishing neutron momentum is a quantity of particular interest.
In order to evaluate $|{\cal M}|^2$ at $q_n{=}0$ we fit the results of the
Fig.~\ref{fig} with a polynomial in $q_n^2$ as suggested in Ref.~\cite{Friman}
\begin{equation}
|{\cal M}|^2 = a+bq_n^2+cq_n{}^4.
\label{fripara}
\end{equation}
We find for the corrected matrix element $a=2.0 \ \times \ 10^3$, $b=-0.6$ fm$^2$ and $c=-0.3$ fm$^4$
and therefore
\begin{equation}
\lim_{q_n \to 0} |{\cal M}|^2 \  = \  (2.0 \pm 0.4) \ \times \ 10^3 \ .
\label{mval}
\end{equation}
It is reasuring that the different parametrization of Eq. (\ref{matrix1}) lead
to the same value of the matrix element at threshold within the experimental 
accuracy.

It is now straight forward to extract a lower limit for the
imaginary part of the elastic scattering length. Assuming that the 
$\omega{N}{\to}\pi{N}$ reaction channel provides the most significant part
of the inelasticity for the ${\omega}N$ scattering, the value
derived should be a reasonable estimate of the
true imaginary part of the scattering length. Note that
model calculations~\cite{Lykasov,Klingl1} show that  the ${\pi}N$ channel 
indeed dominates the $\omega N$ interaction.

We start with the $S$--matrix describing meson--baryon scattering near the 
$\omega$--meson  production threshold and assume $n$ physical final state
channels to be relevant at this energy. The presentation is given  in
the isospin basis. Let the ${\omega}N$ elastic channel be first, while the ${\pi}N$
is second. The matrix element corresponding to the elastic $\omega N$ channel is
 parametrized as 
\begin{equation}
S_{11}:=\eta \exp (2i\delta),
\end{equation}
with $\eta$ being the inelasticity in the ${\omega}N$ scattering and $\delta$ is
the elastic scattering phase shift. 

%The inelasticity at the nominal $\omega$--production threshold is closely
%related to the imaginary part of the $\omega N$ scattering length as
%\begin{equation} 
%\Im \delta =-\frac{1}{2} \ln (\eta ).
%\end{equation}

The unitarity constraint $SS^\dagger=1$ now translates into the relation
\begin{equation}
1-\eta ^2 = \sum_{i=2}^n (2s_i+1)|S_{1i}|^2 \ ,
\label{unit}
\end{equation}
where the factor $(2s_i+1)$ contains the spin multiplicity of the particular 
channel $i$. Since all the contributions in the sum are positive, Eq.~(\ref{unit}) 
directly leads to 
\begin{equation}
1-\eta ^2 \ge 2|S_{12}|^2 \ .
\label{unitb}
\end{equation}

For s--wave scattering the relation between the $S$--matrix and 
$\cal M$ evaluated previously is given by
\begin{equation}
S_{ij} = \delta_{ij} - \frac{i}{2(2\pi)^3}\sqrt{\frac{\lambda_i\lambda_j} 
{\omega_i E_i \omega_j E_j}}(4\pi){\cal M}_{ij},
\end{equation}
where 
\begin{equation}
\lambda_i = \pi\frac{E_i\omega_i}{\sqrt{s}}p_i, 
\end{equation}
with $p_i$ being the 
on--shell momentum in the cms of the particular system and $E_i$ and $\omega_i$
denoting the baryon and meson energy in that channel, respectively.
Combining the two last equations leads to
\begin{equation}
1-\eta^2 \ge \frac{2}{(4\pi)^2}q_n \frac{q_\pi}{s}|{\cal M}|^2,
\label{mofeta}
\end{equation}
where $q_\pi$ and $q_n$ are the incident pion and final nucleon momenta in the
center of mass system, respectively.

The scatting amplitude $f_{\omega{N}}$ in the ${\omega}N$ channel 
is given by
\begin{equation}
f_{\omega N} = \frac{1}{2q_ni}\left(1-\eta e^{2i\delta }\right), 
\end{equation}
and thus its imaginary part reads
\begin{equation}
\Im f_{\omega N} = -\frac{1}{2q_n}(1-\eta \cos (2\delta )).
\end{equation}

The low momentum behavior of both the inelasticity $(1-\eta)$ as well as 
the phase shift $\delta$ is linear in the momentum. Therefore, close to the 
nominal $\omega$ production threshold we evaluate the imaginary part of the
${\omega}N$ scattering length $\Im a_{\omega N}$ as
\begin{equation}
\Im a_{\omega N} := -\lim_{(q_n \to 0)}
\Im f_{\omega N} \ge \frac{1}{2(4\pi)^2}\frac{q_\pi}{s}|{\cal M}|^2.
\label{imformel}
\end{equation}
Note that this formula agrees with the one given in Ref.~\cite{Friman}.

Finally we deduce for the imaginary part of the spin averanged scattering length
\begin{equation}
\Im  a_{\omega N} \ge (0.24 \pm 0.05) \ \mbox{fm} .
\end{equation}

As can be seen from Fig.~\ref{fig}, the value for the corrected squared
matrix element at threshold changed by more than an order
of magnitude. Since the lower limit for $\Im  a_{\omega N} $
scales linearly with the squared matrix element, 
our bound naturally is significantly stronger than those present in the current
literature. For instance, in Ref.~\cite{Friman} a
value $\Im  a_{\omega N}{\ge}0.02$~fm is given (Note:
in Ref.~\cite{Friman} the scattering length was deduced including
the $\pi^-p$ channel only. To include the $\pi^0n$ isotopic channel
the result of Ref.~\cite{Friman} was multiplied with the isospin factor 3/2
to allow comparsion to our value).

\section{Summary}

Using a Monte Carlo simulation as well as an analytical calculation
we reinvestigated the experimental data of Refs. \cite{Binnie,Keyne,Karami}
Based on circumstantial evidence, we argued, that the formalism derived
correctly in Ref. \cite{Binnie} was improperly applied in the analyses.
A purely
kinematical factor not only corrects this error and removes
the unusual energy dependence resulting from the primary analysis, but also
keeps the data in accordance with the world data set.

Based on the modified data we extracted a value for a lower
bound the imaginary part of the elastic $\omega N$ scattering length. The
new value is larger by an order of magnitude compared than that used
in the literature so far \cite{Friman}.
As should be clear from the discussion in the introduction, a large change
in the imaginary part of the scattering length should have a large effect
on estimates for both the in medium width of the omega as well
as the existence of $\omega$--nucleus bound states.

We would like to stress that we call for a remeasurement of the
$\pi N \to \omega N$ reaction in the kinematics close to the nominal
$\omega$ threshold given the importance of this reaction, as outlined
in the introduction. This remeasurement will not only allow to confirm the claims
of this work but should also  fill the gap in the experimental 
data between 200 and 500 MeV/c final center of mass momenta.

\section{Acknowledgment}

We appreciate usefull discussions with J. Durso, A. Kudruyastev, U. Mosel, K.
Nakayama, G. Penner and C. Wilkin. We also would wish to thank J. Durso for careful
reading of the manuscript.

%%%%%%%%%%%%%%%%%%%%%%%%%%%%%%%%%%%%%%%%%%%%%%%%%%%%%%%%%%%%%%%
%%%                      References
%%%%%%%%%%%%%%%%%%%%%%%%%%%%%%%%%%%%%%%%%%%%%%%%%%%%%%%%%%%%%%%


\begin{thebibliography} {99}
\bibitem{labelex}
Here we introduce the
label $exp$ to make explicit that this is the quantity
measured. We will talk about its interpretation in the next
section.
\end{thebibliography}
\end{document}